\documentclass[
 reprint,
superscriptaddress,
 amsmath,amssymb,
pra,
nofootinbib,
]{revtex4-2}
\usepackage{orcidlink}
\usepackage{graphicx}
\usepackage{dcolumn}
\usepackage{bm}
\usepackage{hyperref}
\usepackage{natbib}

\begin{document}

\preprint{APS/123-QED}

\title{Avalanches in Magnetohydrodynamical simulations}

\author{Henri Lamarre\orcidlink{0009-0001-9950-529X}}

\affiliation{Département de Physique, Université de Montréal, Montréal, Québec H2V 0B3, Canada}

\author{Paul Charbonneau\orcidlink{0000-0003-1618-3924}}

\affiliation{Département de Physique, Université de Montréal, Montréal, Québec H2V 0B3, Canada}

\author{Quentin Noraz\orcidlink{0000-0002-7422-1127}}%

\affiliation{Institute of Theoretical Astrophysics, University of Oslo, PO Box 1029 Blindern, 0315 Oslo, Norway}

\author{Antoine Strugarek\orcidlink{0000-0002-9630-6463}}%
%
\author{Alexis Blaise\orcidlink{0009-0008-2500-4100}}%
%
\author{Allan Sacha Brun\orcidlink{0000-0002-1729-8267}}%
%
\affiliation{Université Paris-Saclay, Université Paris Cité, CEA, CNRS, AIM, 91191,
Gif-sur-Yvette, France}%

\author{Mats Carlsson\orcidlink{0000-0001-9218-3139}}%

\author{Boris Vilhelm Gudiksen\orcidlink{0000-0003-0547-4902}}%
%
\affiliation{Institute of Theoretical Astrophysics, University of Oslo, PO Box 1029 Blindern, 0315 Oslo, Norway}
\affiliation{Rosseland Centre for Solar Physics, University of Oslo, P.O. Box 1029 Blindern, 0315 Oslo, Norway}

\date{\today}

\begin{abstract}
Scale invariance is a hallmark of many natural systems, including solar flares, where energy release spans a vast range of scales. Recent computational advances, at the level of both algorithmics and hardware, have enabled high-resolution magnetohydrodynamical (MHD) simulations to span multiple scales, offering new insights into magnetic energy dissipation processes. Here, we study scale invariance of magnetic energy dissipation in two distinct MHD simulations. Current sheets are identified and analyzed over time. Results demonstrate that dissipative events exhibit scale invariance, with power-law distributions characterizing their energy dissipation and lifetimes. Remarkably, these distributions are consistent across the two simulations, despite differing numerical and physical setups, suggesting universality in the process of magnetic energy dissipation. Comparisons between the evolution of dissipation regions reveals distinct growth behaviors in high plasma-$\beta$ regions (convective zone) and low plasma-$\beta$ regions (atmosphere). The latter display spatiotemporal dynamics similar to those of avalanche models, suggesting self-organized criticality and a common universality class.
\end{abstract}

\keywords{}
\maketitle

Scale invariance is found in many natural systems, from the evolution of forest fires to the distribution of earthquake magnitudes \cite{de2006universality}. In the Sun, this property emerges for solar flares, where energy is released across a vast range of scales \cite{aschwanden2000time, shimizu1995energetics, crosby1993frequency}. The most energetic of these events can have significant impacts on Earth \cite{abda2020review} and as such, understanding them is key to produce predictions. However, solar flares have been hard to model from first principles due to the wide disparity of spatial and temporal scales at play. Furthermore, the small-scale nature of magnetic reconnection dynamics, and the complex behaviour of the resulting magnetic energy dissipation \cite{faerder2023comparative, faerder2024comparative} makes the modelling even more arduous.\\ 
Avalanche models have been successful in reproducing the statistical properties of flares \cite{charbonneau2001avalanche, strugarek2014deterministically, thibeault2022forecasting, fuentes2016nanoflare}, and other scale invariant processes, from earthquakes to forest fires \cite{paczuski1996universality, turcotte2004landslides}. In the case of solar flares, these models might be capturing the intricate collective dynamics of reconnection events \cite{chitta2025magnetic, Reid2023avalanches, cozzo2023coronal}. In fact, a nanoflare \cite{parker1988nanoflares} can be conceptually linked to the idea of a local instability.\\
Bridging the gap between scale invariant models for solar flares and comprehensive magnetohydrodynamics (MHD) simulations is nevertheless lacking. Despite capturing the global statistics \cite{charbonneau2001avalanche}, avalanche models rely on ad hoc evolutionary rules, in contrast to the detailed physics described by MHD; reconciling them into a common physical framework is far from trivial \cite{Lu1995PRL, isliker1998solar, liu2002continuum, lamarre2024energy, Reid2023avalanches}. However, MHD simulations can provide physical grounding to these ad hoc rules. State-of-the-art MHD simulations can now resolve finer spatial scales, providing insights into energy dissipation mechanisms that were previously unattainable \cite{faerder2023comparative, faerder2024comparative}. It is now possible to investigate whether energy dissipation in simulated low plasma-$\beta$ environments ($\beta$ is the ratio of gas pressure to magnetic pressure \cite{priest2014magnetohydrodynamics}), is scale invariant, and if it occurs through avalanches.\\
\begin{figure}
    \centering
    \includegraphics[width=7cm]{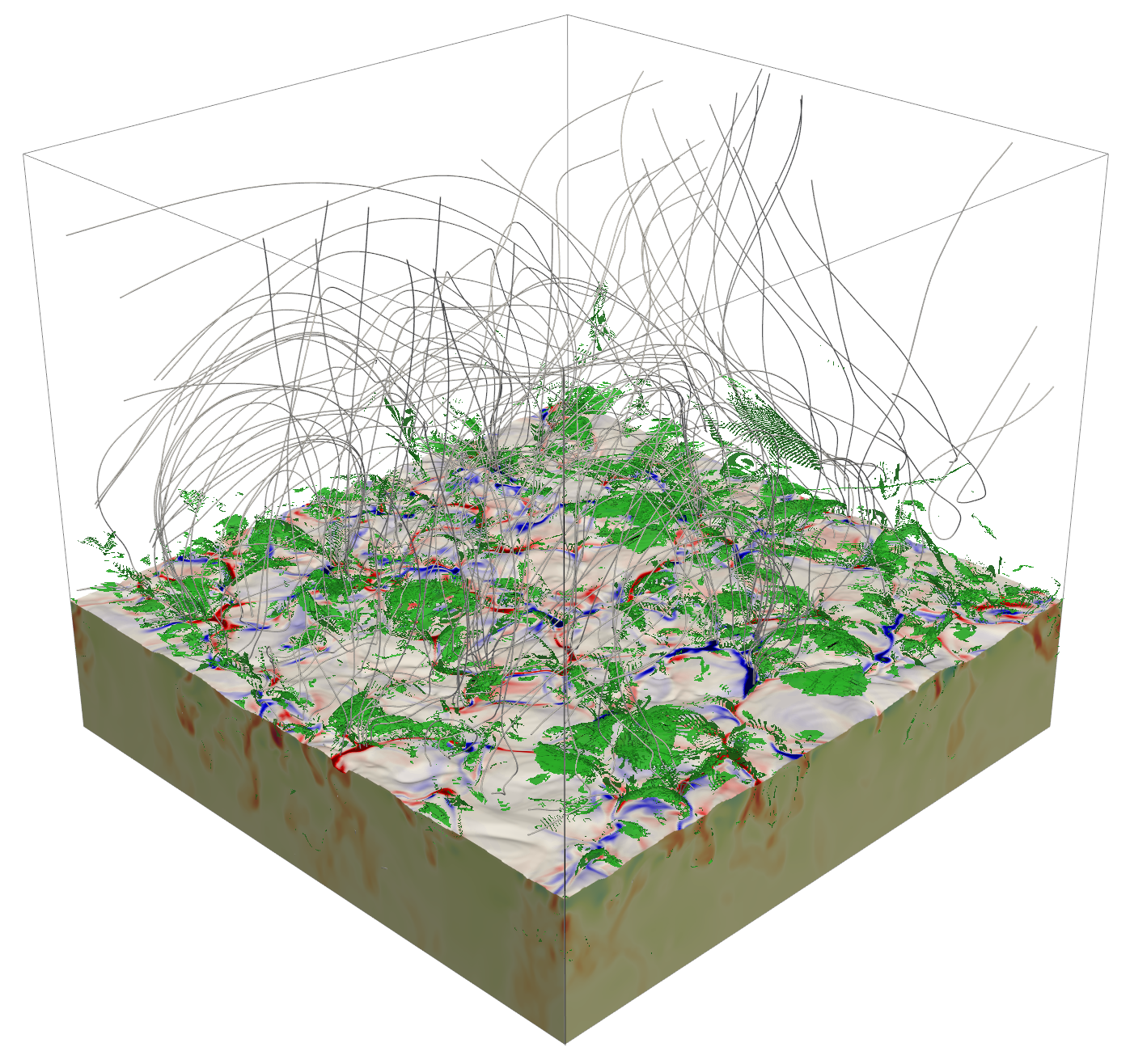}
    \includegraphics[width=7cm]{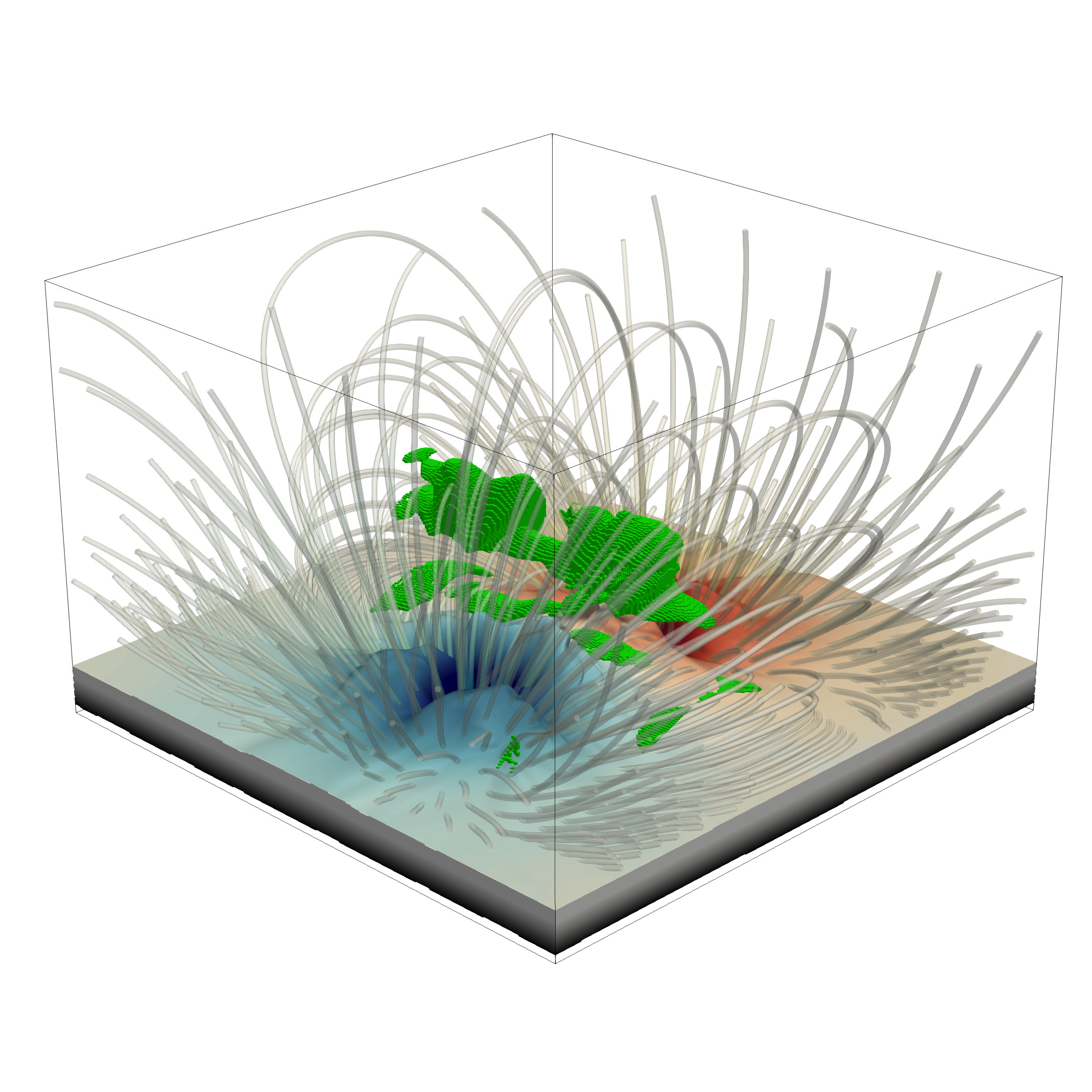} 
    \caption{[Top] The ch012023\_by800 Bifrost simulation. Magnetic field lines in the atmosphere are color coded with the Poynting flux. The photosphere is color coded with $B_{z}$ and the convective zone with vertical convective velocity. [Bottom] PLUTO simulation of a magnetic dipole initially in equilibrium with a stratified atmosphere. The plasma is then rotated at the foot of the dipole. In both panels, green isosurfaces enclose active zones (see text).}
    \label{fig:Sims}
\end{figure}

\begin{figure}
    \centering
    \includegraphics[width=\linewidth]{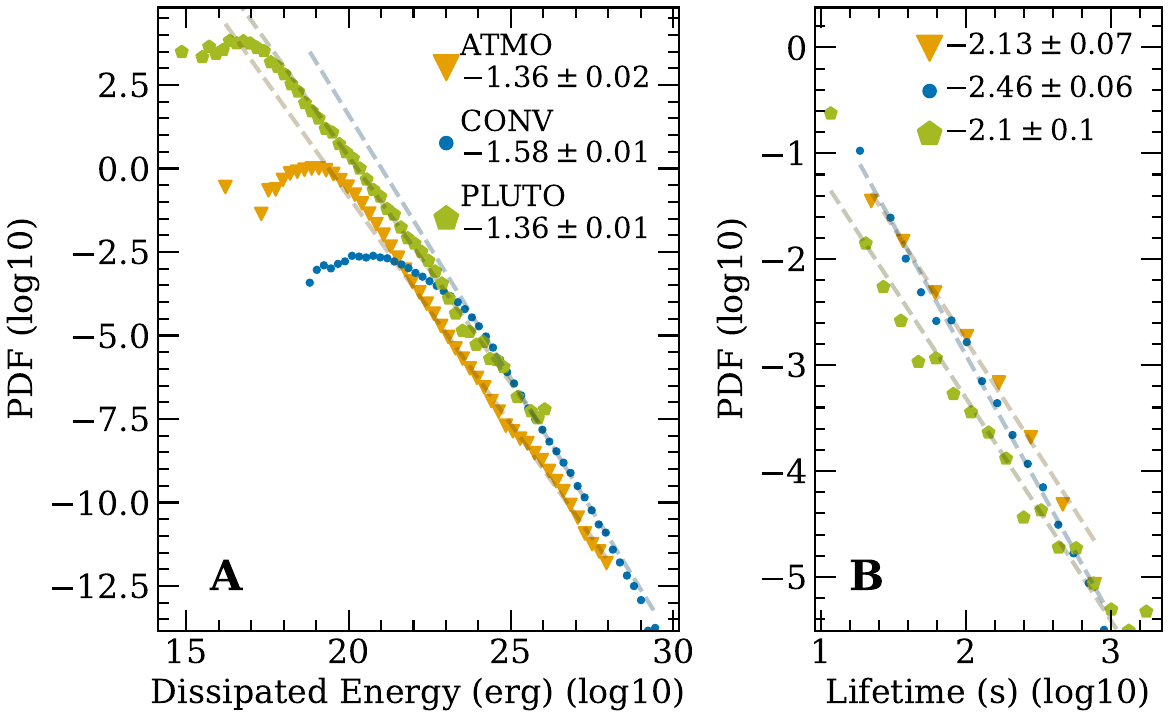}
    \caption{Probability density function of the dissipated magnetic energy (A) and lifetime (B) of the active zones in the Bifrost simulation (ATMO and CONV) and the PLUTO simulation. The cited exponents and uncertainties are recovered with a random sample consensus method \cite{chum2003locally}.}
    \label{fig:Powerlaw}
\end{figure}

To address this question, we analyze magnetic energy dissipation in two distinct MHD simulations. More precisely, we identify regions of high magnetic gradients where ohmic dissipation is expected to occur and analyze their evolution. The two simulation setups we use in this study are displayed in figure \ref{fig:Sims}. The first (top panel) is a state of the art Bifrost simulation \cite{gudiksen2011stellar}, which is a 3D MHD solar-like atmosphere extending from the top of the convective zone to the lower-corona \footnote{Q. Noraz, et al. A\&A (submitted)}. It is a quiet-Sun simulation (ch012023\_by800) experiencing a network-like flux emergence episode. This simulation is designed to capture two dynamically different regimes. First, the convective zone along with the photosphere and lower chromosphere, hereafter CONV, span approximately the lower third of the simulation and are in a high plasma-$\beta$ regime. The upper-chromosphere, the transition region and the beginning of the corona, hereafter ATMO, span the upper two thirds of the simulation and are mostly in a low plasma-$\beta$ regime. We also study a completely different setup in a PLUTO simulation \footnote{A. Blaise, et al. in Prep}, shown in figure \ref{fig:Sims} (bottom panel). This simulation, designed as a test case to study dissipative events, is a bipolar arcade embedded in a stratified low-$\beta$ atmosphere and slowly twisted at its base. 
Saying that these two simulations are different is an understatement. Table \ref{tab:2sims} lists various simulation characteristics and parameter values, to highlight the fact they they differ markedly from one another, in terms of both numerical and physical setups.\\
\begin{table}
    \centering
    \begin{tabular}{|p{2cm}|p{2.4cm}|p{3.6cm}|}\hline
       & PLUTO & Bifrost \\\hline
      Magnetic Topology & Twisted bipolar arcade & Network-like flux emergence in the Quiet-Sun \\\hline
      Dissipative Scheme & Global constant resistivity and augmented local resistivity (YS-94 \cite{faerder2024comparative}) & Explicit operator split into a weak global term and a localized hyperdiffusive term \cite{gudiksen2011stellar} \\\hline
      Plasma-$\beta$ range & $10^{-4}$ - $10^{2}$ & ATMO: $10^{-3}-10^{5}$ \newline CONV: $10^{-2}-10^{7}$ \\\hline
      Average Plasma-$\beta$& $10^{-1}$ & ATMO: $10^{-1}$ \newline CONV: $10^{3}$ \\\hline
      Resolution & 256x256x512 & 512x512x512 \\\hline
      Cell dimension & 70km & Horizontal: 20km\newline Vertical: 12-70km \\\hline
    \end{tabular}
    \caption{Characteristics of the MHD simulations studied in this work.}
    \label{tab:2sims}
\end{table}
We focus on locations with strong magnetic dissipation, which can be interpreted as current sheets. They reveal valuable insights about transfers between the magnetic energy and the internal energy. These locations are identified via the criterion:
\begin{equation}\label{eq:quantity}
    \frac{\|\nabla \times \boldsymbol{B}\|}{\|\boldsymbol{B}\|} > \frac{1}{\Delta},
\end{equation}
where $\Delta$ corresponds to a length scale\footnote{Note that equation \ref{eq:quantity} allows for cases where null points or points where $\|B\|$ is small are tagged as active cells. These points, however, only represent a small ($<0.7\%$) subset of the active cells. In addition, those cells don't dissipate enough magnetic energy to impact the foregoing analysis.}. A low threshold would yield only few very large zones, whereas a high threshold produces multiple small disjointed zones. The value of $\Delta = 0.01$Mm (Bifrost) and $\Delta = 0.035$Mm (PLUTO) which corresponds to $\Delta = \frac{1}{2}$ computational cell width is chosen to maximize simultaneously the number of identified zones and the range of their volumes. However, our results are robust to variations of $\Delta$, as discussed in Appendix \ref{app:delta}.\\
All cells for which the inequality \ref{eq:quantity} holds are defined as active cells. Active zones (hereafter zones) are defined using the Hoshen-Kopelman algorithm \cite{morales2020energy, hoshen1976percolation} which identifies clusters defined as diagonally and horizontally connected active cells, displayed as green isosurfaces in figure \ref{fig:Sims}.
Each zone is assigned a label and an ancestry algorithm\footnote{Our ancestry algorithm is analogous to the one described in \cite{zhdankin2015temporal}. However, in mergers and splitting, which account for 10\% of events, the largest zone keeps its label whereas the smallest is considered ended, or created.} is applied to track their temporal evolution.\\
A first step in characterizing these dissipation zones is to calculate the joule heating 
\begin{equation}\label{eq:heating}
    Q_{J}= 4\pi\eta \frac{{\bf J}^{2}}{c^{2}}
\end{equation} where $\eta$ is the magnetic diffusivity and ${\bf J}$ is the electric current density. It quantifies the transfer of energy from the magnetic field to internal energy. We compute the amount of magnetic energy dissipated by a zone by integrating $Q_{J}$ over its time-varying volume, over its lifetime. From this, we construct probability density functions (PDFs) for the dissipated magnetic energy and lifetime of the zones, as shown in figure \ref{fig:Powerlaw}. These frequency distributions are characterized by extended power-law tails, spanning more than eight orders of magnitudes in energy and two orders of magnitude in lifetimes. The best-fit exponents are listed in the figure. These values are roughly consistent with previous analyzes carried out on a similar Bifrost simulation, using a completely distinct zone identification scheme \cite{kanella2018investigating}, which strengthens our confidence in our identification algorithm.  For all distributions, the low energies generally associated to zones comprised of a few cells are discarded from the fit performed with a random sample consensus method \cite{chum2003locally}. It is noteworthy that the low plasma-$\beta$ regimes of the two simulations, with completely different setups, are described by similar power laws in energies and lifetimes. This result not only confirms that dissipative events are scale invariant in MHD simulations, but also hints at the universality of this process.\\
Many physical mechanisms can lead to the existence of power-laws for the distribution of energy dissipated in localized regions. A paramount example can be found in MHD turbulence (akin to the CONV region in our Bifrost simulation), which is well-known for producing such power-laws \cite{zhdankin2015temporal, boffetta1999power}. However, other generators of power laws exist in physical systems \cite{Aschwanden_2024, sornette2006critical}, such as for instance avalanching processes and criticality. Asserting which frameworks can be used to describe this dissipation thus requires going beyond power-laws.

Avalanching systems have a characteristic evolution \cite{odor2004universality} controlled by a set of local rules of interaction between the smallest components of the system. Scale invariance and universality emerge naturally from such systems. A large event is simply a collection of smaller ones. Under slow driving, many such systems evolve autonomously towards a self-organized critical state, which naturally leads to scale invariance.\\
The Lu and Hamilton \cite{lu1993solar} avalanche model for solar flares (hereafter LH93) is an archetypal example. It is akin to a classical sandpile model \cite{bak1987self}, in which sand grains are added until the local slope exceeds a threshold and the pile becomes unstable. At this point, an unstable site will redistribute some sand to its neighbour to restore stability, but doing so may render its neighbours unstable and trigger another redistribution, which can lead to an avalanche. In this model, scale invariance of avalanches emerges naturally from the stability and redistribution rules, which are purely local, and remains robust to changes in model parameters such as the magnitude of the driving \cite{strugarek2014deterministically}, provided a good separation of timescales exists between energy loading and avalanching.\\
The local interactions will then determine the way unstable regions grow in these systems. This is typically assessed by critical exponents, which quantify the spatiotemporal evolution of these systems and define universality classes \cite{odor2004universality}. In this study, we focus on a spreading exponent \cite{morales2020energy, munoz1999avalanche, morales2008scaling} which characterizes the rate at which unstable zones grow over the lattice. It is recovered from the relation 
\begin{equation}\label{eq:n}
    N(t) = t^{n},
\end{equation}
where $N$ is the number of new active cells at lifetime $t$, averaged over all dissipation events occurring over the duration of a given simulation, and $n$ is the spreading exponent.\\
\begin{figure}
    \centering
    \includegraphics[width=\linewidth]{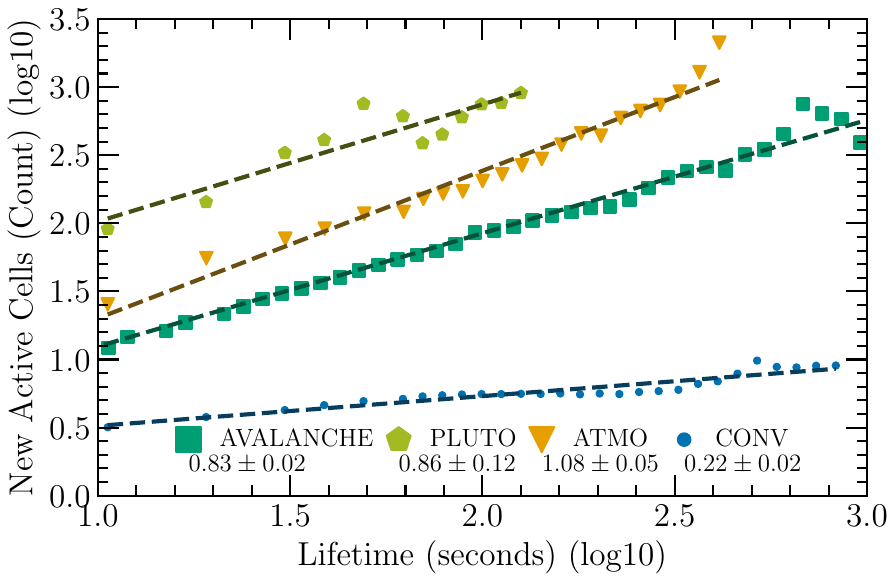}
    \caption{Evolution of the average number of active cells in dissipative zones in the Bifrost simulation, (ATMO and CONV), the PLUTO simulation, and in a Lu and Hamilton 3D model (AVALANCHE). We only retain averages with more than 50 zone samples for statistical significance ($\log_{10}(\text{Lifetime}) < 2$, for PLUTO for example).}
    \label{fig:Spreading}
\end{figure}
We compute the spreading exponent $n$ for CONV and ATMO in the Bifrost simulation, the PLUTO simulation, and a Lu and Hamilton avalanche model (LH93). As shown in figure \ref{fig:Spreading} in all cases, the average number of new active cells $N(t)$ grows as a power-law of time. These are fitted in logspace to measure their slope, directly yielding the spreading exponent $n$.
Zones in the CONV region start on average with $\sim 3$ ($10^{0.5}$) active cells and the number of newly added cells per successive snapshot never exceeds $10$. In contrast, zones in the ATMO region start on average with $\sim 25$ ($10^{1.4}$) active cells and the number of newly added active cells exhibits marked power-law growth. Similar evolution is observed in the PLUTO simulation and the avalanche model\footnote{Time in the LH93 is measured in discrete iterations, but was rescaled to second here for comparison purposes in figure \ref{fig:Spreading}, which does not affect the value of the exponent}. The spreading exponent that quantifies these evolutions are listed in figure \ref{fig:Spreading} as well as in table \ref{tab:spreading}, appendix \ref{app:crit}. Videos of the evolution of sample zones in ATMO and CONV are provided in supplementary material. In those videos, orange cubes are new active cells and green cubes are persisting active cells. In CONV, the new active cells show marked spatiotemporal variability and are vertically aligned, which is to be expected from a high plasma-$\beta$ turbulent regime dominated by convective downflows. In ATMO, the new active cells are typically contiguous to active neighbours and the zone forms a sheet-like structure.\\
To summarize, externally driven homogeneous isotropic turbulence and SOC both generate scale invariance and both dissipate energy at the smallest length scales. However, in turbulence, energy is transfered from large scales to small scales, whereas the opposite can be argued for SOC where small-scale instabilities can destabilize larger areas through avalanches. In that regard, distinguishing between generators requires characterizing the spatiotemporal evolution of energy dissipation regions. In light of this, we assert that the evolution of zones in low plasma-$\beta$ regimes (ATMO and PLUTO), are compatible with avalanche models, and confirm that zones in high plasma-$\beta$ environments, such as CONV evolve differently than in low plasma-$\beta$ environments. The much smaller value of $n$ in CONV further underlines the distinct nature of the dissipative dynamics in the MHD-convective turbulent regime, where we do not expect avalanches to occur. \\ 

\begin{figure}
    \centering
    \includegraphics[width=\linewidth]{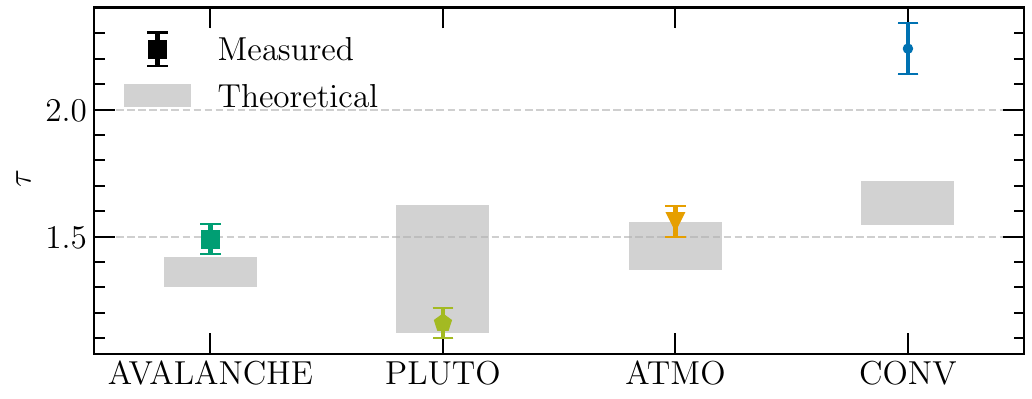}
    \caption{Consistency test for the measured value of $\tau$ (Eq. \ref{eq:tau}) and its theoretical value $\tau^\ast$ (Eq. $\ref{eq:tauStar}$) where a $\pm 2\sigma$ interval is used for the error bars. The larger range on $\tau^{*}$ (gray box) for PLUTO is caused by smaller sample size for zones in that simulation.}
    \label{fig:CritExpos}
\end{figure}

Systems near criticality have characteristic relations between their critical exponents \cite{munoz1999avalanche}. These relations should hold for any system at or near criticality \cite{munoz1999avalanche, morales2008scaling}. Thus, we verify if one such relation holds for our models. This relation, detailed in appendix \ref{app:crit}, is based on another critical exponent, $\tau$, describing the probability distribution for zone sizes. Figure \ref{fig:CritExpos} shows that the low-$\beta$ regimes, ATMO and PLUTO, along with the LH93 model (AVALANCHE) all yield measured values for $\tau$ which are in acceptable agreement with the theoretical value. On the other hand, the measured value of $\tau$ for high plasma-$\beta$ (CONV) is in marked disagreement with the theoretical value $\tau^\ast$, showing that the CONV region is likely far from a critical regime. These reinforce the stark contrast between the high and low $\beta$ regimes in the Bifrost simulation and supporting the conjecture that the associated power-law reflect the turbulent state of the plasma, and does not indicate criticality.\\
The good agreement between the measured and computed critical exponents $\tau$ and $\tau^\ast$ (viz. Fig 4) in the low-$\beta$ region of the Bifrost and PLUTO simulations is indicative of these simulations operating at or near a critical point.
Since neither simulation is subjected to fine tuning via an external control parameter, this criticality must be of the self-organized variety. Self-organized criticality (SOC)\cite{jensen1998self, aschwanden2011self} is understood to materialize in slowly-driven open dissipative systems
subjected to a local threshold instability 
\cite{jensen1998self,Lu1995flares}.
In the Bifrost simulation, the evolution of the ATMO region is driven by the magnetic footpoints being anchored in a turbulent convective layer with typical turnover timescale $\sim 10^3\,$s; whereas in PLUTO a slow global twist is imposed at the footpoint of the pre-existing magnetic arcade, with timescale $\simeq 10^5\,$s. This is much longer than the lifetime of all but the longest dissipation events observed in both simulations (viz. Fig. 2B), so that both systems can be legitimately considered slowly-driven. Activation of enhanced dissipation represents a form of threshold
instability \cite{faerder2023comparative, faerder2024comparative}, which affect plasma conditions locally and eventually brings
the system back below threshold. The basic requirements for SOC are thus satisfied.
Our analyzes thus support the conjecture that the low-$\beta$ regions of the solar chromosphere and lower corona are in a self-organized critical state, with conversion of magnetic energy to heat taking place through avalanches of small dissipative events \cite{parker1988nanoflares}, and scale-free avalanches mapping naturally on scale-free energy release in flares \cite{Lu1995flares}.

\begin{acknowledgments}
This research has been supported by the NSERC Discovery grant RGPIN-2024-04050, the Centre de Recherche en Astrophysique du Québec (CRAQ)S, the European Research Council through the Synergy Grant number 810218 (``The Whole Sun'', ERC-2018-SyG), the Centre National d’Etudes Spatiales (CNES) Solar Orbiter, the Institut National des Sciences de l’Univers (INSU) via the Programme National Soleil-Terre (PNST), the French Agence Nationale de la Recherche (ANR) project STORMGENESIS \#ANR-22-CE31-0013-01, the Institut d'Astrophysique Spatiale (IAS) by the Research Council of Norway through its Centres of Excellence, project number 262622, and Programme for Supercomputing, and by computing HPC and storage re-
sources by GENCI thanks to the grant 2024-A0160410133.\\

\end{acknowledgments}

\appendix
\section{Threshold Variations}\label{app:delta}
\begin{figure}
    \centering
    \includegraphics[width=\linewidth]{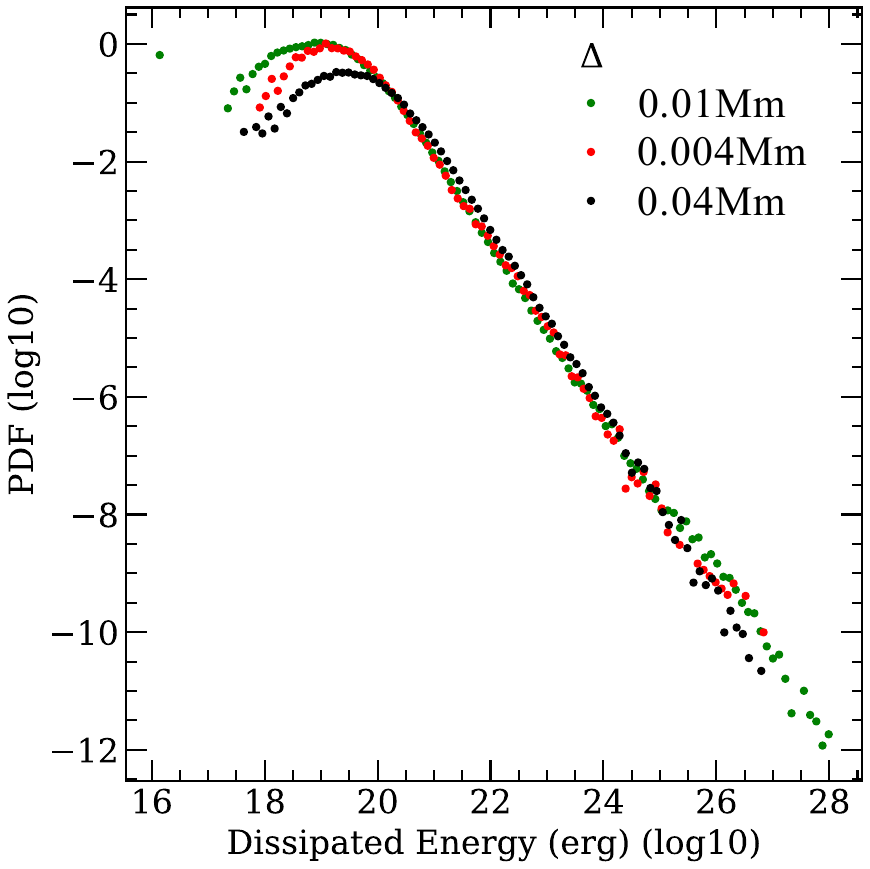}
    \caption{Variation of the distribution of the magnetic energy dissipation as a function of the threshold length scale $\Delta$ used to identify the studied dissipation zones. The results are shown for $\Delta =$ 0.01 Mm (0.5 cells) in green,  0.004 Mm (0.2 cells) in red and 0.04 Mm (2 cells) in black}. The power-law behaviour is essentially unaffected by this choice and $\Delta=$ 0.01 Mm (0.5 cells) is used in this study.
    \label{fig:PL_Delta}
\end{figure}

We have verified that the choice of the activation threshold length scale $\Delta$ in Eq. \ref{eq:delta} does not affect significantly the power-law form and logarithmic slopes of the distribution of dissipation event sizes. As can be seen on Figure \ref{fig:PL_Delta}, only the small event end of the distribution is affected by variations of $\Delta$, while the power-law tail of large events, of primary interest in our study, remains unaffected over more than six orders of magnitude in dissipated energy.

\section{Critical Exponents}\label{app:crit}
Critical exponents characterize the spatiotemporal evolution of systems near their critical point, and can be used collectively to define universality classes \cite{odor2004universality}. These exponents can be recovered by fitting associated distributions of event size measures and are related to one another via theoretically-derived numerical relationships \cite{munoz1999avalanche}. One such exponent, $\tau$, quantifies the growth of an unstable region via the power-law relationship: 
\begin{equation}\label{eq:tau}
    P(S)\sim S^{-\tau},
\end{equation}
where $P(S)$ is the probability that an active zone reaches a size $S$ in its lifetime. This exponent is directly accessible from simulation output, and in critical systems can be related to other critical exponents \cite{munoz1999avalanche, morales2008scaling}: 
\begin{equation}\label{eq:tauStar}
    \tau^{*} =  \frac{1+n+2\delta}{1+n+\delta},
\end{equation}
where $n$ is the spreading exponent defined in equation \ref{eq:n}, and $\delta$ characterizes the probability distribution $P(t)$ that a zone is still active at time $t$:
\begin{equation}\label{eq:delta}
    P(t)\sim t^{-\delta}.
\end{equation}
The three critical exponents $n$, $\delta$ and $\tau$ are fitted on distributions constructed from simulation output, using the random consensus method \cite{chum2003locally}. Their numerical values are listed in table \ref{tab:spreading} along with the value of $\tau^\ast$ computed from Eq.~\ref{eq:tauStar}. These are the data used
to generate Fig.~\ref{fig:CritExpos}.
\begin{table}[!h]
    \begin{tabular}{|p{1.5cm}|p{1.5cm}|p{1.6cm}|p{1.6cm}|p{1.6cm}|}\hline
         Critical Exponent & $n$ - Eq.\ref{eq:n} & $\delta$ - Eq.\ref{eq:delta} & $\tau$ - Eq.\ref{eq:tau} & $\tau^{*}$ - Eq. \ref{eq:tauStar}\\\hline
        ATMO (Bifrost) & $1.08 \pm 0.05$ & $1.8\pm0.1$ & $1.56\pm0.03$ & $1.46\pm0.05$\\\hline
        CONV (Bifrost) & $0.22 \pm 0.02$ & $2.09\pm0.08$ & $2.24\pm0.05$ & $1.63\pm0.04$\\\hline
        PLUTO & $0.9\pm0.1$ & $1.1\pm0.2$  & $1.16\pm0.03$ & $1.4\pm0.1$ \\\hline
        Avalanche Model (LH93) & $0.83 \pm 0.02$ & $1.04\pm0.05$&$1.49\pm0.03$& $1.36\pm0.03$\\\hline
    \end{tabular}
    \caption{Measured and theoretical values for critical exponents in MHD simulations and an avalanche model.}
    \label{tab:spreading}
\end{table}
\providecommand{\noopsort}[1]{}\providecommand{\singleletter}[1]{#1}%

\bibliography{Manuscript}

\providecommand{\noopsort}[1]{}\providecommand{\singleletter}[1]{#1}%
\begin{thebibliography}{39}%
\makeatletter
\providecommand \@ifxundefined [1]{%
 \@ifx{#1\undefined}
}%
\providecommand \@ifnum [1]{%
 \ifnum #1\expandafter \@firstoftwo
 \else \expandafter \@secondoftwo
 \fi
}%
\providecommand \@ifx [1]{%
 \ifx #1\expandafter \@firstoftwo
 \else \expandafter \@secondoftwo
 \fi
}%
\providecommand \natexlab [1]{#1}%
\providecommand \enquote  [1]{``#1''}%
\providecommand \bibnamefont  [1]{#1}%
\providecommand \bibfnamefont [1]{#1}%
\providecommand \citenamefont [1]{#1}%
\providecommand \href@noop [0]{\@secondoftwo}%
\providecommand \href [0]{\begingroup \@sanitize@url \@href}%
\providecommand \@href[1]{\@@startlink{#1}\@@href}%
\providecommand \@@href[1]{\endgroup#1\@@endlink}%
\providecommand \@sanitize@url [0]{\catcode `\\12\catcode `\$12\catcode `\&12\catcode `\#12\catcode `\^12\catcode `\_12\catcode `\%12\relax}%
\providecommand \@@startlink[1]{}%
\providecommand \@@endlink[0]{}%
\providecommand \url  [0]{\begingroup\@sanitize@url \@url }%
\providecommand \@url [1]{\endgroup\@href {#1}{\urlprefix }}%
\providecommand \urlprefix  [0]{URL }%
\providecommand \Eprint [0]{\href }%
\providecommand \doibase [0]{https://doi.org/}%
\providecommand \selectlanguage [0]{\@gobble}%
\providecommand \bibinfo  [0]{\@secondoftwo}%
\providecommand \bibfield  [0]{\@secondoftwo}%
\providecommand \translation [1]{[#1]}%
\providecommand \BibitemOpen [0]{}%
\providecommand \bibitemStop [0]{}%
\providecommand \bibitemNoStop [0]{.\EOS\space}%
\providecommand \EOS [0]{\spacefactor3000\relax}%
\providecommand \BibitemShut  [1]{\csname bibitem#1\endcsname}%
\let\auto@bib@innerbib\@empty
\bibitem [{\citenamefont {de~Arcangelis}\ \emph {et~al.}(2006)\citenamefont {de~Arcangelis}, \citenamefont {Godano}, \citenamefont {Lippiello},\ and\ \citenamefont {Nicodemi}}]{de2006universality}%
  \BibitemOpen
  \bibfield  {author} {\bibinfo {author} {\bibfnamefont {L.}~\bibnamefont {de~Arcangelis}}, \bibinfo {author} {\bibfnamefont {C.}~\bibnamefont {Godano}}, \bibinfo {author} {\bibfnamefont {E.}~\bibnamefont {Lippiello}},\ and\ \bibinfo {author} {\bibfnamefont {M.}~\bibnamefont {Nicodemi}},\ }\bibfield  {title} {\bibinfo {title} {Universality in solar flare and earthquake occurrence},\ }\href@noop {} {\bibfield  {journal} {\bibinfo  {journal} {Physical review letters}\ }\textbf {\bibinfo {volume} {96}},\ \bibinfo {pages} {051102} (\bibinfo {year} {2006})}\BibitemShut {NoStop}%
\bibitem [{\citenamefont {Aschwanden}\ \emph {et~al.}(2000)\citenamefont {Aschwanden}, \citenamefont {Tarbell}, \citenamefont {Nightingale}, \citenamefont {Schrijver}, \citenamefont {Kankelborg}, \citenamefont {Martens}, \citenamefont {Warren} \emph {et~al.}}]{aschwanden2000time}%
  \BibitemOpen
  \bibfield  {author} {\bibinfo {author} {\bibfnamefont {M.~J.}\ \bibnamefont {Aschwanden}}, \bibinfo {author} {\bibfnamefont {T.~D.}\ \bibnamefont {Tarbell}}, \bibinfo {author} {\bibfnamefont {R.~W.}\ \bibnamefont {Nightingale}}, \bibinfo {author} {\bibfnamefont {C.~J.}\ \bibnamefont {Schrijver}}, \bibinfo {author} {\bibfnamefont {C.~C.}\ \bibnamefont {Kankelborg}}, \bibinfo {author} {\bibfnamefont {P.}~\bibnamefont {Martens}}, \bibinfo {author} {\bibfnamefont {H.~P.}\ \bibnamefont {Warren}}, \emph {et~al.},\ }\bibfield  {title} {\bibinfo {title} {Time variability of the “quiet” sun observed with trace. ii. physical parameters, temperature evolution, and energetics of extreme-ultraviolet nanoflares},\ }\href@noop {} {\bibfield  {journal} {\bibinfo  {journal} {The Astrophysical Journal}\ }\textbf {\bibinfo {volume} {535}},\ \bibinfo {pages} {1047} (\bibinfo {year} {2000})}\BibitemShut {NoStop}%
\bibitem [{\citenamefont {Shimizu}(1995)}]{shimizu1995energetics}%
  \BibitemOpen
  \bibfield  {author} {\bibinfo {author} {\bibfnamefont {T.}~\bibnamefont {Shimizu}},\ }\bibfield  {title} {\bibinfo {title} {Energetics and occurrence rate of active-region transient brightenings and implications for the heating of the active-region corona},\ }\href@noop {} {\bibfield  {journal} {\bibinfo  {journal} {Publications of the Astronomical Society of Japan, v. 47, p. 251-263.}\ }\textbf {\bibinfo {volume} {47}},\ \bibinfo {pages} {251} (\bibinfo {year} {1995})}\BibitemShut {NoStop}%
\bibitem [{\citenamefont {Crosby}\ \emph {et~al.}(1993)\citenamefont {Crosby}, \citenamefont {Aschwanden},\ and\ \citenamefont {Dennis}}]{crosby1993frequency}%
  \BibitemOpen
  \bibfield  {author} {\bibinfo {author} {\bibfnamefont {N.~B.}\ \bibnamefont {Crosby}}, \bibinfo {author} {\bibfnamefont {M.~J.}\ \bibnamefont {Aschwanden}},\ and\ \bibinfo {author} {\bibfnamefont {B.~R.}\ \bibnamefont {Dennis}},\ }\bibfield  {title} {\bibinfo {title} {Frequency distributions and correlations of solar x-ray flare parameters},\ }\href@noop {} {\bibfield  {journal} {\bibinfo  {journal} {Solar Physics}\ }\textbf {\bibinfo {volume} {143}},\ \bibinfo {pages} {275} (\bibinfo {year} {1993})}\BibitemShut {NoStop}%
\bibitem [{\citenamefont {Abda}\ \emph {et~al.}(2020)\citenamefont {Abda}, \citenamefont {Ab~Aziz}, \citenamefont {Ab~Kadir},\ and\ \citenamefont {Rhazali}}]{abda2020review}%
  \BibitemOpen
  \bibfield  {author} {\bibinfo {author} {\bibfnamefont {Z.~M.~K.}\ \bibnamefont {Abda}}, \bibinfo {author} {\bibfnamefont {N.~F.}\ \bibnamefont {Ab~Aziz}}, \bibinfo {author} {\bibfnamefont {M.~Z.~A.}\ \bibnamefont {Ab~Kadir}},\ and\ \bibinfo {author} {\bibfnamefont {Z.~A.}\ \bibnamefont {Rhazali}},\ }\bibfield  {title} {\bibinfo {title} {A review of geomagnetically induced current effects on electrical power system: Principles and theory},\ }\href@noop {} {\bibfield  {journal} {\bibinfo  {journal} {IEEE Access}\ }\textbf {\bibinfo {volume} {8}},\ \bibinfo {pages} {200237} (\bibinfo {year} {2020})}\BibitemShut {NoStop}%
\bibitem [{\citenamefont {F{\ae}rder}\ \emph {et~al.}(2023)\citenamefont {F{\ae}rder}, \citenamefont {N{\'o}brega-Siverio},\ and\ \citenamefont {Carlsson}}]{faerder2023comparative}%
  \BibitemOpen
  \bibfield  {author} {\bibinfo {author} {\bibfnamefont {{\O}.~H.}\ \bibnamefont {F{\ae}rder}}, \bibinfo {author} {\bibfnamefont {D.}~\bibnamefont {N{\'o}brega-Siverio}},\ and\ \bibinfo {author} {\bibfnamefont {M.}~\bibnamefont {Carlsson}},\ }\bibfield  {title} {\bibinfo {title} {A comparative study of resistivity models for simulations of magnetic reconnection in the solar atmosphere},\ }\href@noop {} {\bibfield  {journal} {\bibinfo  {journal} {Astronomy \& Astrophysics}\ }\textbf {\bibinfo {volume} {675}},\ \bibinfo {pages} {A97} (\bibinfo {year} {2023})}\BibitemShut {NoStop}%
\bibitem [{\citenamefont {F{\ae}rder}\ \emph {et~al.}(2024)\citenamefont {F{\ae}rder}, \citenamefont {N{\'o}brega-Siverio},\ and\ \citenamefont {Carlsson}}]{faerder2024comparative}%
  \BibitemOpen
  \bibfield  {author} {\bibinfo {author} {\bibfnamefont {{\O}.~H.}\ \bibnamefont {F{\ae}rder}}, \bibinfo {author} {\bibfnamefont {D.}~\bibnamefont {N{\'o}brega-Siverio}},\ and\ \bibinfo {author} {\bibfnamefont {M.}~\bibnamefont {Carlsson}},\ }\bibfield  {title} {\bibinfo {title} {A comparative study of resistivity models for simulations of magnetic reconnection in the solar atmosphere-ii. plasmoid formation},\ }\href@noop {} {\bibfield  {journal} {\bibinfo  {journal} {Astronomy \& Astrophysics}\ }\textbf {\bibinfo {volume} {683}},\ \bibinfo {pages} {A95} (\bibinfo {year} {2024})}\BibitemShut {NoStop}%
\bibitem [{\citenamefont {Charbonneau}\ \emph {et~al.}(2001)\citenamefont {Charbonneau}, \citenamefont {McIntosh}, \citenamefont {Liu},\ and\ \citenamefont {Bogdan}}]{charbonneau2001avalanche}%
  \BibitemOpen
  \bibfield  {author} {\bibinfo {author} {\bibfnamefont {P.}~\bibnamefont {Charbonneau}}, \bibinfo {author} {\bibfnamefont {S.~W.}\ \bibnamefont {McIntosh}}, \bibinfo {author} {\bibfnamefont {H.-L.}\ \bibnamefont {Liu}},\ and\ \bibinfo {author} {\bibfnamefont {T.~J.}\ \bibnamefont {Bogdan}},\ }\bibfield  {title} {\bibinfo {title} {Avalanche models for solar flares (invited review)},\ }\href@noop {} {\bibfield  {journal} {\bibinfo  {journal} {Solar Physics}\ }\textbf {\bibinfo {volume} {203}},\ \bibinfo {pages} {321} (\bibinfo {year} {2001})}\BibitemShut {NoStop}%
\bibitem [{\citenamefont {Strugarek}\ \emph {et~al.}(2014)\citenamefont {Strugarek}, \citenamefont {Charbonneau}, \citenamefont {Joseph},\ and\ \citenamefont {Pirot}}]{strugarek2014deterministically}%
  \BibitemOpen
  \bibfield  {author} {\bibinfo {author} {\bibfnamefont {A.}~\bibnamefont {Strugarek}}, \bibinfo {author} {\bibfnamefont {P.}~\bibnamefont {Charbonneau}}, \bibinfo {author} {\bibfnamefont {R.}~\bibnamefont {Joseph}},\ and\ \bibinfo {author} {\bibfnamefont {D.}~\bibnamefont {Pirot}},\ }\bibfield  {title} {\bibinfo {title} {Deterministically driven avalanche models of solar flares},\ }\href@noop {} {\bibfield  {journal} {\bibinfo  {journal} {Coronal Magnetometry}\ ,\ \bibinfo {pages} {371}} (\bibinfo {year} {2014})}\BibitemShut {NoStop}%
\bibitem [{\citenamefont {Thibeault}\ \emph {et~al.}(2022)\citenamefont {Thibeault}, \citenamefont {Strugarek}, \citenamefont {Charbonneau},\ and\ \citenamefont {Tremblay}}]{thibeault2022forecasting}%
  \BibitemOpen
  \bibfield  {author} {\bibinfo {author} {\bibfnamefont {C.}~\bibnamefont {Thibeault}}, \bibinfo {author} {\bibfnamefont {A.}~\bibnamefont {Strugarek}}, \bibinfo {author} {\bibfnamefont {P.}~\bibnamefont {Charbonneau}},\ and\ \bibinfo {author} {\bibfnamefont {B.}~\bibnamefont {Tremblay}},\ }\bibfield  {title} {\bibinfo {title} {Forecasting solar flares by data assimilation in sandpile models. solar physics},\ }\href@noop {} {\bibfield  {journal} {\bibinfo  {journal} {arXiv preprint arXiv:2206.13583}\ } (\bibinfo {year} {2022})}\BibitemShut {NoStop}%
\bibitem [{\citenamefont {Fuentes}\ and\ \citenamefont {Klimchuk}(2016)}]{fuentes2016nanoflare}%
  \BibitemOpen
  \bibfield  {author} {\bibinfo {author} {\bibfnamefont {M.~L.}\ \bibnamefont {Fuentes}}\ and\ \bibinfo {author} {\bibfnamefont {J.~A.}\ \bibnamefont {Klimchuk}},\ }\bibfield  {title} {\bibinfo {title} {A nanoflare-based cellular automaton model and the observed properties of the coronal plasma},\ }\href@noop {} {\bibfield  {journal} {\bibinfo  {journal} {The Astrophysical Journal}\ }\textbf {\bibinfo {volume} {828}},\ \bibinfo {pages} {86} (\bibinfo {year} {2016})}\BibitemShut {NoStop}%
\bibitem [{\citenamefont {Paczuski}\ and\ \citenamefont {Boettcher}(1996)}]{paczuski1996universality}%
  \BibitemOpen
  \bibfield  {author} {\bibinfo {author} {\bibfnamefont {M.}~\bibnamefont {Paczuski}}\ and\ \bibinfo {author} {\bibfnamefont {S.}~\bibnamefont {Boettcher}},\ }\bibfield  {title} {\bibinfo {title} {Universality in sandpiles, interface depinning, and earthquake models},\ }\href@noop {} {\bibfield  {journal} {\bibinfo  {journal} {Physical review letters}\ }\textbf {\bibinfo {volume} {77}},\ \bibinfo {pages} {111} (\bibinfo {year} {1996})}\BibitemShut {NoStop}%
\bibitem [{\citenamefont {Turcotte}\ and\ \citenamefont {Malamud}(2004)}]{turcotte2004landslides}%
  \BibitemOpen
  \bibfield  {author} {\bibinfo {author} {\bibfnamefont {D.~L.}\ \bibnamefont {Turcotte}}\ and\ \bibinfo {author} {\bibfnamefont {B.~D.}\ \bibnamefont {Malamud}},\ }\bibfield  {title} {\bibinfo {title} {Landslides, forest fires, and earthquakes: examples of self-organized critical behavior},\ }\href@noop {} {\bibfield  {journal} {\bibinfo  {journal} {Physica A: Statistical Mechanics and its Applications}\ }\textbf {\bibinfo {volume} {340}},\ \bibinfo {pages} {580} (\bibinfo {year} {2004})}\BibitemShut {NoStop}%
\bibitem [{\citenamefont {Chitta}\ \emph {et~al.}(2025)\citenamefont {Chitta}, \citenamefont {Pontin}, \citenamefont {Priest}, \citenamefont {Berghmans}, \citenamefont {Kraaikamp}, \citenamefont {Rodriguez}, \citenamefont {Verbeeck}, \citenamefont {Zhukov}, \citenamefont {Krucker}, \citenamefont {Cuadrado} \emph {et~al.}}]{chitta2025magnetic}%
  \BibitemOpen
  \bibfield  {author} {\bibinfo {author} {\bibfnamefont {L.}~\bibnamefont {Chitta}}, \bibinfo {author} {\bibfnamefont {D.}~\bibnamefont {Pontin}}, \bibinfo {author} {\bibfnamefont {E.}~\bibnamefont {Priest}}, \bibinfo {author} {\bibfnamefont {D.}~\bibnamefont {Berghmans}}, \bibinfo {author} {\bibfnamefont {E.}~\bibnamefont {Kraaikamp}}, \bibinfo {author} {\bibfnamefont {L.}~\bibnamefont {Rodriguez}}, \bibinfo {author} {\bibfnamefont {C.}~\bibnamefont {Verbeeck}}, \bibinfo {author} {\bibfnamefont {A.}~\bibnamefont {Zhukov}}, \bibinfo {author} {\bibfnamefont {S.}~\bibnamefont {Krucker}}, \bibinfo {author} {\bibfnamefont {R.~A.}\ \bibnamefont {Cuadrado}}, \emph {et~al.},\ }\bibfield  {title} {\bibinfo {title} {A magnetic avalanche as the central engine powering a solar flare},\ }\href@noop {} {\bibfield  {journal} {\bibinfo  {journal} {arXiv preprint arXiv:2503.12235}\ } (\bibinfo {year} {2025})}\BibitemShut {NoStop}%
\bibitem [{\citenamefont {{Reid}}\ \emph {et~al.}(2023)\citenamefont {{Reid}}, \citenamefont {{Threlfall}},\ and\ \citenamefont {{Hood}}}]{Reid2023avalanches}%
  \BibitemOpen
  \bibfield  {author} {\bibinfo {author} {\bibfnamefont {J.}~\bibnamefont {{Reid}}}, \bibinfo {author} {\bibfnamefont {J.}~\bibnamefont {{Threlfall}}},\ and\ \bibinfo {author} {\bibfnamefont {A.~W.}\ \bibnamefont {{Hood}}},\ }\bibfield  {title} {\bibinfo {title} {{Self-consistent nanoflare heating in model active regions: MHD avalanches}},\ }\href {https://doi.org/10.1093/mnras/stac3188} {\bibfield  {journal} {\bibinfo  {journal} {MNRAS}\ }\textbf {\bibinfo {volume} {518}},\ \bibinfo {pages} {1584} (\bibinfo {year} {2023})}\BibitemShut {NoStop}%
\bibitem [{\citenamefont {Cozzo}\ \emph {et~al.}(2023)\citenamefont {Cozzo}, \citenamefont {Reid}, \citenamefont {Pagano}, \citenamefont {Reale},\ and\ \citenamefont {Hood}}]{cozzo2023coronal}%
  \BibitemOpen
  \bibfield  {author} {\bibinfo {author} {\bibfnamefont {G.}~\bibnamefont {Cozzo}}, \bibinfo {author} {\bibfnamefont {J.}~\bibnamefont {Reid}}, \bibinfo {author} {\bibfnamefont {P.}~\bibnamefont {Pagano}}, \bibinfo {author} {\bibfnamefont {F.}~\bibnamefont {Reale}},\ and\ \bibinfo {author} {\bibfnamefont {A.~W.}\ \bibnamefont {Hood}},\ }\bibfield  {title} {\bibinfo {title} {Coronal energy release by mhd avalanches-effects on a structured, active region, multi-threaded coronal loop},\ }\href@noop {} {\bibfield  {journal} {\bibinfo  {journal} {Astronomy \& Astrophysics}\ }\textbf {\bibinfo {volume} {678}},\ \bibinfo {pages} {A40} (\bibinfo {year} {2023})}\BibitemShut {NoStop}%
\bibitem [{\citenamefont {Parker}(1988)}]{parker1988nanoflares}%
  \BibitemOpen
  \bibfield  {author} {\bibinfo {author} {\bibfnamefont {E.~N.}\ \bibnamefont {Parker}},\ }\bibfield  {title} {\bibinfo {title} {Nanoflares and the solar x-ray corona},\ }\href@noop {} {\bibfield  {journal} {\bibinfo  {journal} {Astrophysical Journal, Part 1 (ISSN 0004-637X), vol. 330, July 1, 1988, p. 474-479.}\ }\textbf {\bibinfo {volume} {330}},\ \bibinfo {pages} {474} (\bibinfo {year} {1988})}\BibitemShut {NoStop}%
\bibitem [{\citenamefont {{Lu}}(1995{\natexlab{a}})}]{Lu1995PRL}%
  \BibitemOpen
  \bibfield  {author} {\bibinfo {author} {\bibfnamefont {E.~T.}\ \bibnamefont {{Lu}}},\ }\bibfield  {title} {\bibinfo {title} {{Avalanches in Continuum Driven Dissipative Systems}},\ }\href {https://doi.org/10.1103/PhysRevLett.74.2511} {\bibfield  {journal} {\bibinfo  {journal} {\prl}\ }\textbf {\bibinfo {volume} {74}},\ \bibinfo {pages} {2511} (\bibinfo {year} {1995}{\natexlab{a}})}\BibitemShut {NoStop}%
\bibitem [{\citenamefont {Isliker}\ \emph {et~al.}(1998)\citenamefont {Isliker}, \citenamefont {Anastasiadis}, \citenamefont {Vassiliadis},\ and\ \citenamefont {Vlahos}}]{isliker1998solar}%
  \BibitemOpen
  \bibfield  {author} {\bibinfo {author} {\bibfnamefont {H.}~\bibnamefont {Isliker}}, \bibinfo {author} {\bibfnamefont {A.}~\bibnamefont {Anastasiadis}}, \bibinfo {author} {\bibfnamefont {D.}~\bibnamefont {Vassiliadis}},\ and\ \bibinfo {author} {\bibfnamefont {L.}~\bibnamefont {Vlahos}},\ }\bibfield  {title} {\bibinfo {title} {Solar flare cellular automata interpreted as discretized mhd equations},\ }\href@noop {} {\bibfield  {journal} {\bibinfo  {journal} {Astronomy and Astrophysics, v. 335, p. 1085-1092 (1998)}\ }\textbf {\bibinfo {volume} {335}},\ \bibinfo {pages} {1085} (\bibinfo {year} {1998})}\BibitemShut {NoStop}%
\bibitem [{\citenamefont {Liu}\ \emph {et~al.}(2002)\citenamefont {Liu}, \citenamefont {Charbonneau}, \citenamefont {Pouquet}, \citenamefont {Bogdan},\ and\ \citenamefont {McIntosh}}]{liu2002continuum}%
  \BibitemOpen
  \bibfield  {author} {\bibinfo {author} {\bibfnamefont {H.-L.}\ \bibnamefont {Liu}}, \bibinfo {author} {\bibfnamefont {P.}~\bibnamefont {Charbonneau}}, \bibinfo {author} {\bibfnamefont {A.}~\bibnamefont {Pouquet}}, \bibinfo {author} {\bibfnamefont {T.}~\bibnamefont {Bogdan}},\ and\ \bibinfo {author} {\bibfnamefont {S.}~\bibnamefont {McIntosh}},\ }\bibfield  {title} {\bibinfo {title} {Continuum analysis of an avalanche model for solar flares},\ }\href@noop {} {\bibfield  {journal} {\bibinfo  {journal} {Physical Review E}\ }\textbf {\bibinfo {volume} {66}},\ \bibinfo {pages} {056111} (\bibinfo {year} {2002})}\BibitemShut {NoStop}%
\bibitem [{\citenamefont {Lamarre}\ \emph {et~al.}(2024)\citenamefont {Lamarre}, \citenamefont {Charbonneau},\ and\ \citenamefont {Strugarek}}]{lamarre2024energy}%
  \BibitemOpen
  \bibfield  {author} {\bibinfo {author} {\bibfnamefont {H.}~\bibnamefont {Lamarre}}, \bibinfo {author} {\bibfnamefont {P.}~\bibnamefont {Charbonneau}},\ and\ \bibinfo {author} {\bibfnamefont {A.}~\bibnamefont {Strugarek}},\ }\bibfield  {title} {\bibinfo {title} {Energy definition and minimization in avalanche models for solar flares},\ }\href@noop {} {\bibfield  {journal} {\bibinfo  {journal} {Solar Physics}\ }\textbf {\bibinfo {volume} {299}},\ \bibinfo {pages} {13} (\bibinfo {year} {2024})}\BibitemShut {NoStop}%
\bibitem [{\citenamefont {Priest}(2014)}]{priest2014magnetohydrodynamics}%
  \BibitemOpen
  \bibfield  {author} {\bibinfo {author} {\bibfnamefont {E.}~\bibnamefont {Priest}},\ }\href@noop {} {\emph {\bibinfo {title} {Magnetohydrodynamics of the Sun}}}\ (\bibinfo  {publisher} {Cambridge University Press},\ \bibinfo {year} {2014})\BibitemShut {NoStop}%
\bibitem [{\citenamefont {Gudiksen}\ \emph {et~al.}(2011)\citenamefont {Gudiksen}, \citenamefont {Carlsson}, \citenamefont {Hansteen}, \citenamefont {Hayek}, \citenamefont {Leenaarts},\ and\ \citenamefont {Mart{\'\i}nez-Sykora}}]{gudiksen2011stellar}%
  \BibitemOpen
  \bibfield  {author} {\bibinfo {author} {\bibfnamefont {B.~V.}\ \bibnamefont {Gudiksen}}, \bibinfo {author} {\bibfnamefont {M.}~\bibnamefont {Carlsson}}, \bibinfo {author} {\bibfnamefont {V.~H.}\ \bibnamefont {Hansteen}}, \bibinfo {author} {\bibfnamefont {W.}~\bibnamefont {Hayek}}, \bibinfo {author} {\bibfnamefont {J.}~\bibnamefont {Leenaarts}},\ and\ \bibinfo {author} {\bibfnamefont {J.}~\bibnamefont {Mart{\'\i}nez-Sykora}},\ }\bibfield  {title} {\bibinfo {title} {The stellar atmosphere simulation code bifrost-code description and validation},\ }\href@noop {} {\bibfield  {journal} {\bibinfo  {journal} {Astronomy \& Astrophysics}\ }\textbf {\bibinfo {volume} {531}},\ \bibinfo {pages} {A154} (\bibinfo {year} {2011})}\BibitemShut {NoStop}%
\bibitem [{\citenamefont {Morales}\ \emph {et~al.}(2020)\citenamefont {Morales}, \citenamefont {Dmitruk},\ and\ \citenamefont {G{\'o}mez}}]{morales2020energy}%
  \BibitemOpen
  \bibfield  {author} {\bibinfo {author} {\bibfnamefont {L.~F.}\ \bibnamefont {Morales}}, \bibinfo {author} {\bibfnamefont {P.}~\bibnamefont {Dmitruk}},\ and\ \bibinfo {author} {\bibfnamefont {D.~O.}\ \bibnamefont {G{\'o}mez}},\ }\bibfield  {title} {\bibinfo {title} {Energy dissipation in coronal loops: Statistical analysis of intermittent structures in magnetohydrodynamic turbulence},\ }\href@noop {} {\bibfield  {journal} {\bibinfo  {journal} {The Astrophysical Journal}\ }\textbf {\bibinfo {volume} {894}},\ \bibinfo {pages} {90} (\bibinfo {year} {2020})}\BibitemShut {NoStop}%
\bibitem [{\citenamefont {Hoshen}\ and\ \citenamefont {Kopelman}(1976)}]{hoshen1976percolation}%
  \BibitemOpen
  \bibfield  {author} {\bibinfo {author} {\bibfnamefont {J.}~\bibnamefont {Hoshen}}\ and\ \bibinfo {author} {\bibfnamefont {R.}~\bibnamefont {Kopelman}},\ }\bibfield  {title} {\bibinfo {title} {Percolation and cluster distribution. i. cluster multiple labeling technique and critical concentration algorithm},\ }\href@noop {} {\bibfield  {journal} {\bibinfo  {journal} {Physical Review B}\ }\textbf {\bibinfo {volume} {14}},\ \bibinfo {pages} {3438} (\bibinfo {year} {1976})}\BibitemShut {NoStop}%
\bibitem [{\citenamefont {Zhdankin}\ \emph {et~al.}(2015)\citenamefont {Zhdankin}, \citenamefont {Uzdensky},\ and\ \citenamefont {Boldyrev}}]{zhdankin2015temporal}%
  \BibitemOpen
  \bibfield  {author} {\bibinfo {author} {\bibfnamefont {V.}~\bibnamefont {Zhdankin}}, \bibinfo {author} {\bibfnamefont {D.~A.}\ \bibnamefont {Uzdensky}},\ and\ \bibinfo {author} {\bibfnamefont {S.}~\bibnamefont {Boldyrev}},\ }\bibfield  {title} {\bibinfo {title} {Temporal analysis of dissipative structures in magnetohydrodynamic turbulence},\ }\href@noop {} {\bibfield  {journal} {\bibinfo  {journal} {The Astrophysical Journal}\ }\textbf {\bibinfo {volume} {811}},\ \bibinfo {pages} {6} (\bibinfo {year} {2015})}\BibitemShut {NoStop}%
\bibitem [{\citenamefont {Kanella}\ and\ \citenamefont {Gudiksen}(2018)}]{kanella2018investigating}%
  \BibitemOpen
  \bibfield  {author} {\bibinfo {author} {\bibfnamefont {C.}~\bibnamefont {Kanella}}\ and\ \bibinfo {author} {\bibfnamefont {B.~V.}\ \bibnamefont {Gudiksen}},\ }\bibfield  {title} {\bibinfo {title} {Investigating 4d coronal heating events in magnetohydrodynamic simulations},\ }\href@noop {} {\bibfield  {journal} {\bibinfo  {journal} {Astronomy \& Astrophysics}\ }\textbf {\bibinfo {volume} {617}},\ \bibinfo {pages} {A50} (\bibinfo {year} {2018})}\BibitemShut {NoStop}%
\bibitem [{\citenamefont {Chum}\ \emph {et~al.}(2003)\citenamefont {Chum}, \citenamefont {Matas},\ and\ \citenamefont {Kittler}}]{chum2003locally}%
  \BibitemOpen
  \bibfield  {author} {\bibinfo {author} {\bibfnamefont {O.}~\bibnamefont {Chum}}, \bibinfo {author} {\bibfnamefont {J.}~\bibnamefont {Matas}},\ and\ \bibinfo {author} {\bibfnamefont {J.}~\bibnamefont {Kittler}},\ }\bibfield  {title} {\bibinfo {title} {Locally optimized ransac},\ }in\ \href@noop {} {\emph {\bibinfo {booktitle} {Pattern Recognition: 25th DAGM Symposium, Magdeburg, Germany, September 10-12, 2003. Proceedings 25}}}\ (\bibinfo {organization} {Springer},\ \bibinfo {year} {2003})\ pp.\ \bibinfo {pages} {236--243}\BibitemShut {NoStop}%
\bibitem [{\citenamefont {Boffetta}\ \emph {et~al.}(1999)\citenamefont {Boffetta}, \citenamefont {Carbone}, \citenamefont {Giuliani}, \citenamefont {Veltri},\ and\ \citenamefont {Vulpiani}}]{boffetta1999power}%
  \BibitemOpen
  \bibfield  {author} {\bibinfo {author} {\bibfnamefont {G.}~\bibnamefont {Boffetta}}, \bibinfo {author} {\bibfnamefont {V.}~\bibnamefont {Carbone}}, \bibinfo {author} {\bibfnamefont {P.}~\bibnamefont {Giuliani}}, \bibinfo {author} {\bibfnamefont {P.}~\bibnamefont {Veltri}},\ and\ \bibinfo {author} {\bibfnamefont {A.}~\bibnamefont {Vulpiani}},\ }\bibfield  {title} {\bibinfo {title} {Power laws in solar flares: self-organized criticality or turbulence?},\ }\href@noop {} {\bibfield  {journal} {\bibinfo  {journal} {Physical review letters}\ }\textbf {\bibinfo {volume} {83}},\ \bibinfo {pages} {4662} (\bibinfo {year} {1999})}\BibitemShut {NoStop}%
\bibitem [{\citenamefont {Aschwanden}(2024)}]{Aschwanden_2024}%
  \BibitemOpen
  \bibfield  {author} {\bibinfo {author} {\bibfnamefont {M.}~\bibnamefont {Aschwanden}},\ }\href@noop {} {\emph {\bibinfo {title} {Power Laws in Astrophysics: Self-Organized Criticality Systems}}}\ (\bibinfo  {publisher} {Cambridge University Press},\ \bibinfo {year} {2024})\BibitemShut {NoStop}%
\bibitem [{\citenamefont {Sornette}(2006)}]{sornette2006critical}%
  \BibitemOpen
  \bibfield  {author} {\bibinfo {author} {\bibfnamefont {D.}~\bibnamefont {Sornette}},\ }\href@noop {} {\emph {\bibinfo {title} {Critical phenomena in natural sciences: chaos, fractals, selforganization and disorder: concepts and tools}}}\ (\bibinfo  {publisher} {Springer Science \& Business Media},\ \bibinfo {year} {2006})\BibitemShut {NoStop}%
\bibitem [{\citenamefont {{\'O}dor}(2004)}]{odor2004universality}%
  \BibitemOpen
  \bibfield  {author} {\bibinfo {author} {\bibfnamefont {G.}~\bibnamefont {{\'O}dor}},\ }\bibfield  {title} {\bibinfo {title} {Universality classes in nonequilibrium lattice systems},\ }\href@noop {} {\bibfield  {journal} {\bibinfo  {journal} {Reviews of modern physics}\ }\textbf {\bibinfo {volume} {76}},\ \bibinfo {pages} {663} (\bibinfo {year} {2004})}\BibitemShut {NoStop}%
\bibitem [{\citenamefont {Lu}\ \emph {et~al.}(1993)\citenamefont {Lu}, \citenamefont {Hamilton}, \citenamefont {McTiernan},\ and\ \citenamefont {Bromund}}]{lu1993solar}%
  \BibitemOpen
  \bibfield  {author} {\bibinfo {author} {\bibfnamefont {E.~T.}\ \bibnamefont {Lu}}, \bibinfo {author} {\bibfnamefont {R.~J.}\ \bibnamefont {Hamilton}}, \bibinfo {author} {\bibfnamefont {J.}~\bibnamefont {McTiernan}},\ and\ \bibinfo {author} {\bibfnamefont {K.~R.}\ \bibnamefont {Bromund}},\ }\bibfield  {title} {\bibinfo {title} {Solar flares and avalanches in driven dissipative systems},\ }\href@noop {} {\bibfield  {journal} {\bibinfo  {journal} {Astrophysical Journal, Part 1 (ISSN 0004-637X), vol. 412, no. 2, p. 841-852.}\ }\textbf {\bibinfo {volume} {412}},\ \bibinfo {pages} {841} (\bibinfo {year} {1993})}\BibitemShut {NoStop}%
\bibitem [{\citenamefont {Bak}\ \emph {et~al.}(1987)\citenamefont {Bak}, \citenamefont {Tang},\ and\ \citenamefont {Wiesenfeld}}]{bak1987self}%
  \BibitemOpen
  \bibfield  {author} {\bibinfo {author} {\bibfnamefont {P.}~\bibnamefont {Bak}}, \bibinfo {author} {\bibfnamefont {C.}~\bibnamefont {Tang}},\ and\ \bibinfo {author} {\bibfnamefont {K.}~\bibnamefont {Wiesenfeld}},\ }\bibfield  {title} {\bibinfo {title} {Self-organized criticality: An explanation of the 1/f noise},\ }\href@noop {} {\bibfield  {journal} {\bibinfo  {journal} {Physical review letters}\ }\textbf {\bibinfo {volume} {59}},\ \bibinfo {pages} {381} (\bibinfo {year} {1987})}\BibitemShut {NoStop}%
\bibitem [{\citenamefont {Munoz}\ \emph {et~al.}(1999)\citenamefont {Munoz}, \citenamefont {Dickman}, \citenamefont {Vespignani},\ and\ \citenamefont {Zapperi}}]{munoz1999avalanche}%
  \BibitemOpen
  \bibfield  {author} {\bibinfo {author} {\bibfnamefont {M.~A.}\ \bibnamefont {Munoz}}, \bibinfo {author} {\bibfnamefont {R.}~\bibnamefont {Dickman}}, \bibinfo {author} {\bibfnamefont {A.}~\bibnamefont {Vespignani}},\ and\ \bibinfo {author} {\bibfnamefont {S.}~\bibnamefont {Zapperi}},\ }\bibfield  {title} {\bibinfo {title} {Avalanche and spreading exponents in systems with absorbing states},\ }\href@noop {} {\bibfield  {journal} {\bibinfo  {journal} {Physical Review E}\ }\textbf {\bibinfo {volume} {59}},\ \bibinfo {pages} {6175} (\bibinfo {year} {1999})}\BibitemShut {NoStop}%
\bibitem [{\citenamefont {Morales}\ and\ \citenamefont {Charbonneau}(2008)}]{morales2008scaling}%
  \BibitemOpen
  \bibfield  {author} {\bibinfo {author} {\bibfnamefont {L.~F.}\ \bibnamefont {Morales}}\ and\ \bibinfo {author} {\bibfnamefont {P.}~\bibnamefont {Charbonneau}},\ }\bibfield  {title} {\bibinfo {title} {Scaling laws and frequency distributions of avalanche areas in a self-organized criticality model of solar flares},\ }\href@noop {} {\bibfield  {journal} {\bibinfo  {journal} {Geophysical research letters}\ }\textbf {\bibinfo {volume} {35}} (\bibinfo {year} {2008})}\BibitemShut {NoStop}%
\bibitem [{\citenamefont {Jensen}(1998)}]{jensen1998self}%
  \BibitemOpen
  \bibfield  {author} {\bibinfo {author} {\bibfnamefont {H.~J.}\ \bibnamefont {Jensen}},\ }\href@noop {} {\emph {\bibinfo {title} {Self-organized criticality: emergent complex behavior in physical and biological systems}}},\ Vol.~\bibinfo {volume} {10}\ (\bibinfo  {publisher} {Cambridge university press},\ \bibinfo {year} {1998})\BibitemShut {NoStop}%
\bibitem [{\citenamefont {Aschwanden}(2011)}]{aschwanden2011self}%
  \BibitemOpen
  \bibfield  {author} {\bibinfo {author} {\bibfnamefont {M.}~\bibnamefont {Aschwanden}},\ }\href@noop {} {\emph {\bibinfo {title} {Self-organized criticality in astrophysics: The statistics of nonlinear processes in the universe}}}\ (\bibinfo  {publisher} {Springer Science \& Business Media},\ \bibinfo {year} {2011})\BibitemShut {NoStop}%
\bibitem [{\citenamefont {{Lu}}(1995{\natexlab{b}})}]{Lu1995flares}%
  \BibitemOpen
  \bibfield  {author} {\bibinfo {author} {\bibfnamefont {E.~T.}\ \bibnamefont {{Lu}}},\ }\bibfield  {title} {\bibinfo {title} {{The Statistical Physics of Solar Active Regions and the Fundamental Nature of Solar Flares}},\ }\href {https://doi.org/10.1086/187942} {\bibfield  {journal} {\bibinfo  {journal} {APJL}\ }\textbf {\bibinfo {volume} {446}},\ \bibinfo {pages} {L109} (\bibinfo {year} {1995}{\natexlab{b}})}\BibitemShut {NoStop}%
\end{thebibliography}%

\end{document}